\documentclass[preprintnumbers,a4paper,11pt]{revtex4}
\usepackage{amsfonts,amsthm,amsmath,amssymb,graphicx,hyperref,color,subfigure}

\begin{document}
 \bibliographystyle{plain}

\title{Detecting Elastic pp Scattering by Radiative Photons in CMS at the LHC\footnote{Talk given by H.~Gr\"onqvist at XL International Symposium on Multiparticle Dynamics (ISMD) in Antwerp, Belgium, September 2010.}}

\author{H.~Gr\"onqvist$^1$, V.~A.~Khoze$^2$, J.~W.~L\"ams\"a$^{3,4}$, M.~Murray$^{5 \ddagger}$ and R.~Orava$^{1,4 \ddagger}$  \\  
 $^1$ Department of Physics, University of Helsinki, Finland \\
 $^2$ IPPP, Department of Physics, Durham University, Durham DH1 3LE, UK \\
 $^3$ Iowa State University, Ames, Iowa, U.S.A. \\
 $^4$ Helsinki Institute of Physics, PL64, 00014 University of Helsinki, Finland \\
 $^5$ Department of Physics and Astronomy, University of Kansas, U.S.A. \\
 $^\ddagger$ Presently at CERN: CERN-PE, CH-1211 Geneva 23}

\preprint{IPPP/10/96}
\preprint{ DCPT/10/192}

\begin{abstract}
  Photon bremsstrahlung is studied for identifying elastic proton-proton interactions in the CMS experiment at the LHC. In addition to measurement of the elastic pp cross section (assuming that the elastic slope is known) the bremsstrahlung photons will allow evaluation of the total pp cross section, luminosity and alignment of the Zero Degree Calorimeters (ZDCs).
\end{abstract}

\maketitle

\section{Introduction} 
Elastic proton-proton interactions can be tagged at the LHC by detecting the bremsstrahlung photons \cite{Khoze:2010hg}.

\begin{figure}[htb]
\begin{center}
\includegraphics[scale=0.5]{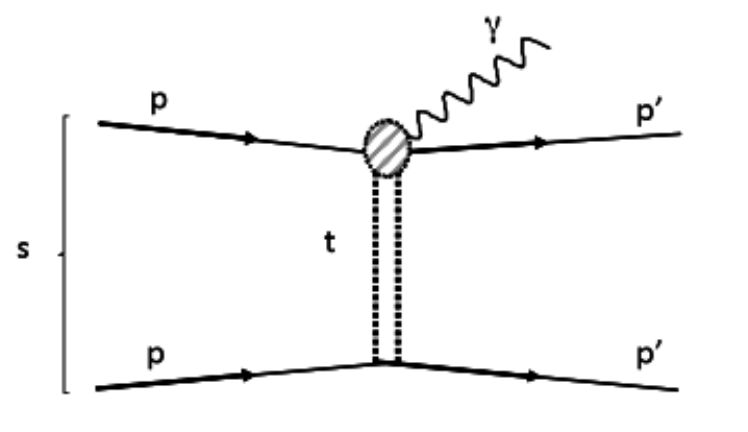}
\caption{Feynman diagram describing the interaction.}
\end{center}
\label{fig:feynman}
\end{figure}
\noindent The probability to radiate a soft photon with energy $k<<E$  ($k\equiv \vert \vec k \vert$) is given by (see for example~ \cite{Berestetskii})

\begin{equation}
\frac{4\pi}{\rho} \frac{ \text{d}\sigma_k }{\sigma^{pp}_{el}}=\frac{\alpha_{em}}{\pi}\:� \text{d}\cos \theta_k \frac{\text{d}k}{k} \int_0^1 \theta_s \text{d}\theta_s \exp({-\rho \frac{\theta_s^2}{2}}) \int_0^{2\pi} \text{d}\phi \:� \mathcal{F} \:Р,  \label{eq:probability}
\end{equation}
where

\begin{equation} \mathcal{F} = \frac{4\theta^2_k}{\left[ \frac{m^2}{E^2}+\theta_k^2 \right]^2}+\frac{4\theta^2_{k^\prime}}{\left[ \frac{m^2}{E^{\prime 2}}+\theta_{k^\prime}^2 \right]^2}-\frac{8\theta_k(\theta_k-\theta_s \cos \phi)}{\left[ \frac{m^2}{E^2}+\theta_k^2 \right]\left[ \frac{m^2}{E^{\prime 2}}+\theta_{k^\prime}^2 \right]} \label{eq:F}
\end{equation} 
and $\rho = B s / 2 $, $\theta_{k}$ is the photon emission angle, $\theta_{k^\prime}$ is the angle between the photon $\vec k$ and the outgoing proton 3-momenta $\vec p~^\prime$, $\theta_s$ is the proton scattering angle, $\phi$ is the azimuthal angle between the photon and the outgoing proton momentum, $E$ and $E^\prime$ the initial and final state proton energy, respectively, and $m$ is the proton rest mass.

The angular distribution of the emitted photons will depend on the slope, B, of the differential cross section $ \text{d}\sigma^{pp}_{el} / {\text{d}t} \longrightarrow \sigma^{pp}_{el}\:�B \exp \left(-B \vert t \vert \right)   $, since $ \vert t \vert \rightarrow p^2\theta^2_s = p^2_t $. 

Characteristically the angle of the radiated photon with respect to the (incident) proton direction, namely $\theta_k = m/ E$, is significantly larger than the proton scattering angle $ \theta_s \approx p_t /E$. This is due to the slope parameter being approximately $ B \approx 20$ GeV$^{-2}$ at LHC energies, giving a mean $\langle p_t \rangle \approx 1/ \sqrt{B} \approx 0.22$ GeV $ \ll m$.  In the limit $ \left( \theta_s / \theta_k \right)^2 \ll 1$ the probability for the emittance of a soft photon with energy $k$ simplifies after integration over the solid angle to~\cite{Khoze:2010hg}

\begin{equation}
\Gamma_\gamma = \frac{2\alpha_{em}}{3 \pi} \frac{ \left\langle p_t^2 \right\rangle }{m^2} \frac{dk}{k} \:� .
\label{eq:GammaGamma}
\end{equation} 
As seen from Equations \eqref{eq:probability}-\eqref{eq:GammaGamma} the cross section for soft photon bremsstrahlung is proportional to $ \sigma(pp)_{el} \left\langle p_t^2 \right\rangle =  \sigma(pp)_{el} / B$. Considering only the real part of the elastic amplitude and using the relation $\sigma_{el} = \sigma_{tot}^2/16\pi B$ the radiative cross section is seen to be proportional to $ \left( \sigma_{el}/\sigma_{tot}\right)^2$.

The ratios $\sigma_{el}/B$ and $\sigma_{el}/ \sigma_{tot}$ do not depend on energy when one assumes geometric scaling~\cite{Buras:1973km}. Geometric scaling implies that all the cross sections and the slope $B$ are proportional to each other and depend only on the interaction radius $R^2(s)$. However, at LHC energies the elastic amplitude should reach the black disk limit in the center of the disk, which violates the geometric scaling hypothesis.

Based on  a compilation of the most recent theoretical results~\cite{Ryskin:2009tj} the ratio $\left( \sigma_{el}/\sigma_{tot} \right)^2$ is in the range $0.0515 \;Ð.\:Ð.\:Ð.\;Ð 0.0876$. This in spite of the fact that at lower energies all the models describe the measured total and elastic cross sections reasonably well. These models are based on the same basic principles and differ only in the way they sum the multi-pomeron vertices. A review of existing theoretical models is given in~\cite{Fiore:2008tp}.

It is worth mentioning the important advantages of measuring the two soft  bremsstrahlung photons emitted in the opposite directions along the incoming beams.
First of all, the coincidence of two photon registration provides better separation of the signal from the backgrounds and triggering conditions. Secondly,  a simultaneous detection of both single and double bremsstrahlung will allow a luminosity independent determination of the effective elastic slope, $B$, and then knowing the luminosity and $B$ we can find the elastic cross section $\sigma_{el}$.

Some time ago double photon bremsstrahlung at small angles in electron collisions,  calculated in \cite{bfk}, was successfully used as a high precision luminosity monitor at electron colliding beam facilities in Novosibirsk \cite{nov} and Orsay \cite{ors} and other laboratories.

\section{Experiment}

\begin{figure}[htb]
\begin{center}
\includegraphics{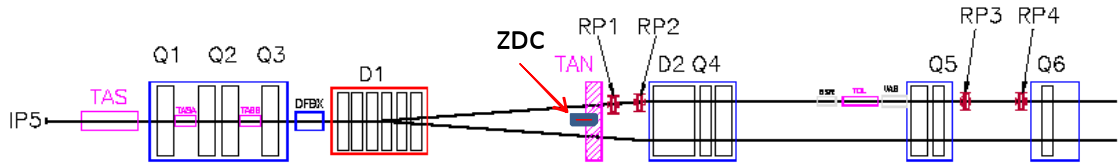}
\caption{The lay-out of the forward region of the CMS Interaction Point 5 (IP5). The locations of the absorbers (TAS and TAN), Quadrupole magnets (Q1-Q6), Dipole magnets (D1, D2) and TOTEM Roman Pot stations (RP1-RP4) at +147 and +220 meters are shown. The CMS Zero Degree Calorimeters (ZDC) are located at 140 meters from IP5, where the LHC vacuum chamber separates in two: The ZDCs only detect neutral secondaries emitted at 0 degrees since the charged beam-like particles will be bent by the D1 dipole magnets~\cite{Grachov1a}.}
\label{fig:experiment} 
\end{center} 
\end{figure}

The relatively soft photons emitted by the incoming and outgoing $5$ TeV protons are registered at the Zero Degree Calorimeters (ZDCs) which are installed at approximately  $\pm 140$ meters from the interaction point (Figure \ref{fig:experiment}).

The CMS experiment is equipped with a ZDC system ~\cite{Grachov1a}. The ZDCs are able to detect very forward photons and neutrons and they measure the particle energy and the transverse position of the particle (in the horizontal direction). The ZDCs are located at approximately zero degrees to the incident beams on each side of the interaction point, beyond the first separation dipole at $\vert z\vert = 140$ m. Having a transverse size of approximately 8 cm $\times$ 8 cm they are designed to fit in between the two beam pipes, covering a rapidity range of $\vert \eta \vert > 8.5$.

Photons within an angular cone of approximately 0.3 mr will strike the ZDC. The physics signal is characterized by a photon distribution that peaks at $r^2=0$. Here $r^2$ is the radial distance squared in the transverse $(x,y)$-plane, as seen from the center of the ZDC. The CMS ZDCs detect transverse distance only in the $x$-direction, integrating over the $y$-range of the detector. Since the electromagnetic section is 19 radiation lengths in the z-direction, total integration over $z$ is assumed.

\section{Background}

The main contribution to the background consists of photons from inelastic diffractive pp interactions, since they are favored by a forward production mechanism. The signal due to these, however, is expected to be uniform over the small area covered by the ZDC, resulting in a distribution that is flat in $r^2$. Photons from non-diffractive events constitute a secondary background.

In order to enhance the signal-to-background ratio we consider a restricted energy range of 50 GeV to 500 GeV. Simulations with PYTHIA ~\cite{pythia} and GEANT~\cite{geant} show that neutron contribution arises above 500 GeV approximately. Photons from $\pi^0$ decay and from particle interactions in the beam pipes are almost entirely in the energy range below 50 GeV. The relative intensity of the background is estimated to be $< 5 \%$ in the chosen energy range.

Background arising from multiple photon hits in the ZDC has been studied in simulations. For the energy range under consideration the ratio of two-photon hits to one-photon hits is around 15$\%$ for the {\emph{T1T2 veto}} condition~\cite{Khoze:2010hg}. For more than one photon hits the energy distribution in the transverse plane is markedly different from the one obtained by considering just single hits. This is caused by the separation of the photons. The multiple hit background is assumed to be significantly reduced and brought down to acceptable levels by shower profile fitting. For details concerning different veto conditions and triggers see~\cite{Khoze:2010hg}.

\section{Results}
\subsection{Simulating the photon distribution}

\begin{figure}[ht]
\begin{minipage}[b]{0.5\linewidth}
\centering
\includegraphics[scale=0.25]{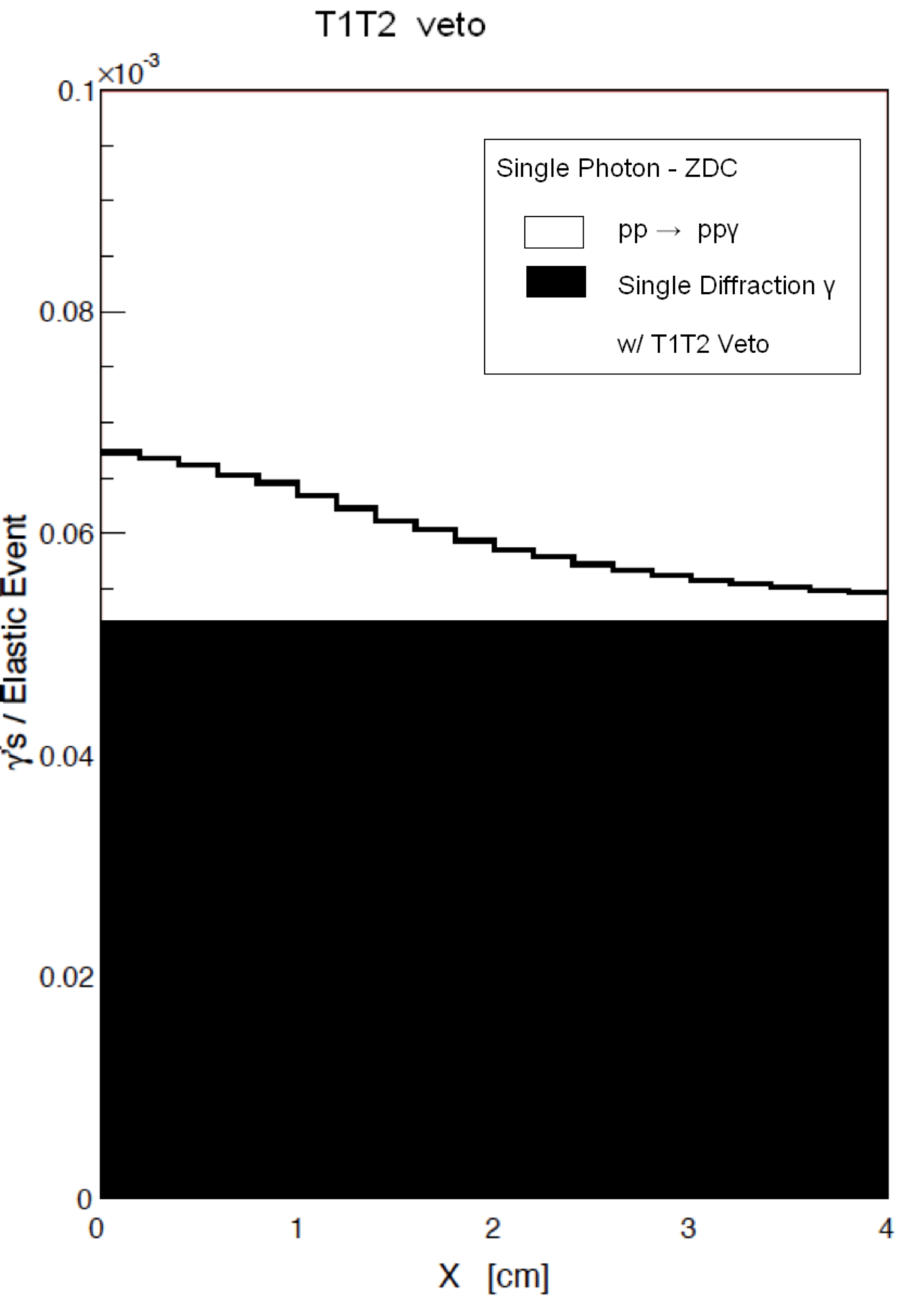}
\end{minipage}
\hspace{0.5cm}
\begin{minipage}[b]{0.5\linewidth}
\centering
\includegraphics[scale=0.25]{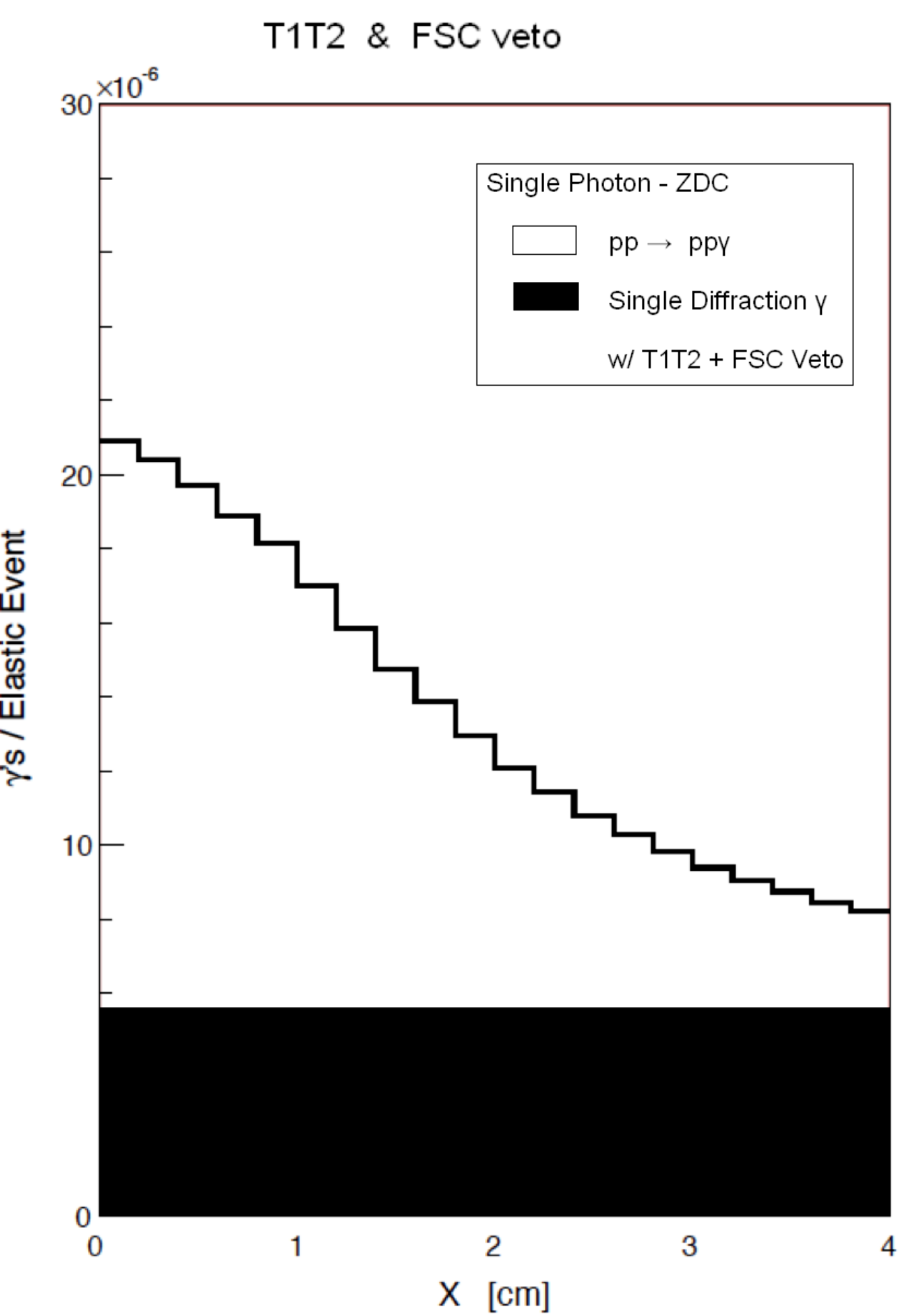}
\end{minipage}
\caption{The probability of detecting a single photon in the ZDC from radiative elastic scattering (open historgam) added together with the probability for detecting a photon from a single diffractive (SD) event (shaded portion) as a function of the horizontal distance, x, of the photon hit from the center of the ZDC. The normalization is with respect to an elastic event. On the top is the case for {\emph{T1T2 veto}}, on the bottom for {\emph{T1T2 + FSC veto}}.}
\label{fig:result1}
\end{figure}

The probability of detecting a single photon in the ZDC from radiative elastic scattering, as a function of x, is shown by the open histograms in Figure \ref{fig:result1}.  The radiative photon contribution is plotted on top of the background distribution for the probability of a singular photon from single-diffractive events to strike the ZDC.  The normalization is taken respect to the cross-section for the production of a single elastic event. Both high-mass and low-mass dissociation are included in the PYTHIA simulation.  Factors are introduced which give an enhancement of the low-mass region, where resonance structure is observed in the data. \\

The calculations assume various trigger conditions, the conditions {\emph{T1T2}} and {\emph{T1T2 + FSC}} are displayed in Figure \ref{fig:result1}. The simulation was also done for the {\emph{no veto}} condition~\cite{Khoze:2010hg}. For the {\emph{T1T2 veto}} condition a distinct signal is seen above the background, and in the case of {\emph{T1T2 + FSC}} veto the signal-to-background ratio is large. 

\subsection{Reconstructing the vertex of photon hits}
Once the photon distribution predicted by simulations is known the next step taken is to consider the performance of the ZDC in a measurement of the type considered. The  electromagnetic section of the ZDC of CMS is divided into five horizontal (transverse) readout towers, with quartz fibres in between absorber plates of tungsten. The tungsten plates are oriented vertically and fibres from the individual towers are bunched together to form a readout tower bundle~\cite{Grachov1a}. Thus the coordinate of a photon hit can be measured only in the $x$ -direction, in the range $ -4<x<  +4$. The width of a tower is approximately 1.6 cm. 

\begin{figure}[htb]
\centering
\includegraphics[scale=0.4]{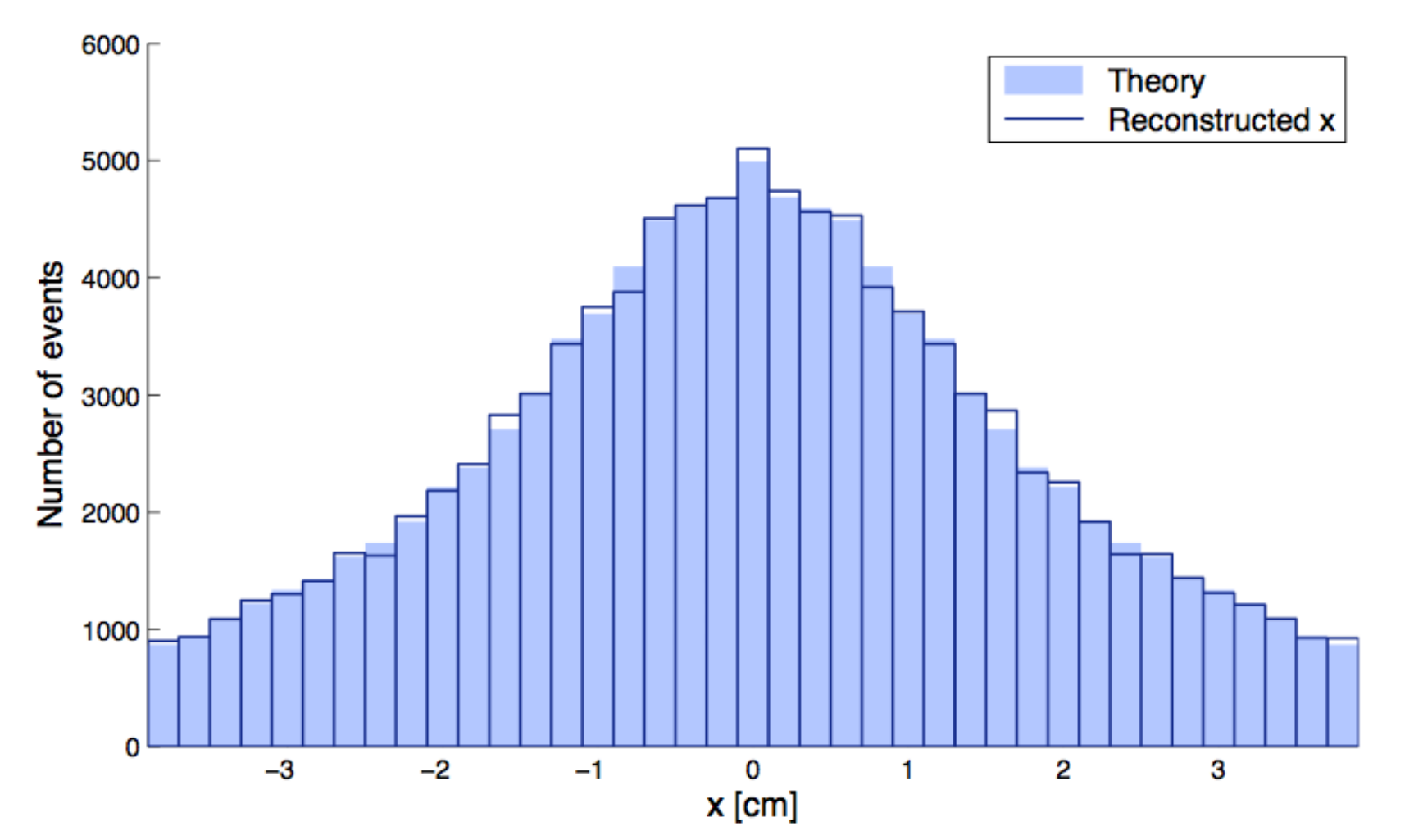}
\caption{Reconstructing the vertex of a photon hit, simulation with $10^5$ events and photon energy 195 GeV.}
\label{fig:matlab}
\end{figure}

The photons that hit the ZDC produce showers of a known width (Moli\`ere radius) of $\sim 0.5$cm. The energy of a shower is spread out transversely in a two-dimensional Gaussian distribution with a standard deviation defined as the Moli\`ere radius of the shower. The ZDC measures the energy deposited in each tower. Photon showers in the ZDC were studied using the photon bremsstrahlung formula.  By examining the ratio of energies in the two adjacent towers with the most energy, the vertices of photon hits can be reconstructed.  These energy ratios (from theory) were created and then used in the vertex reconstruction. Finally, simulated photon hits were produced with energy smearing for each tower, using the expected resolution of the ZDC~\cite{Grachov1a}.

\begin{equation}
\frac{\sigma}{E}= \sqrt{ \left( \frac{0.70}{E}\right)^2 + \left(�0.03�\right)^2}
\end{equation}
The vertex of a given hit is reconstructed by comparing the simulated energy ratios to the ones given by theory.  The result is shown in Figure~\ref{fig:matlab}. 

\section{Conclusions}
By measuring photon bremsstrahlung from protons, elastic pp scattering events can be identified at the LHC. The photons radiated off the initial and final state protons will be seen by the Zero Degree Calorimeter and can be used to measure the product $\sigma^{pp}_{el} \left\langle p_t^2 \right\rangle$ or the ratio $\sigma^{pp}_{el} / \sigma^{pp}_{tot}$~\cite{Khoze:2010hg},~\cite{Fiore:2008tp}, luminosity~\cite{Khoze:2000db} and relative alignment of the ZDCs and of the Roman Pot detectors. The forward detectors covering rapidities $\vert \eta \vert >3$ provide an efficient veto against neutral particle backgrounds in the ZDCs from diffractive and non-diffractive events. The proposed Forward Shower Counters (FSCs) would significantly improve this veto efficiency~\cite{fsc}. Finally, the above analysis should encourage the use of all forward detector systems in LHC experiments, so that the maximum physics discovery potential can be achieved.

\section{Acknowledgements}
The authors are grateful to Albert De Roeck for clarifying discussions and to Misha Ryskin for collaboration and useful discussions. VAK thanks University of Helsinki and Helsinki Institute of Physics for hospitality and HEPTOOLS ITN for support. RO gratefully acknowledges the Academy of Finland for support.

\begin{footnotesize}

\end{footnotesize}


\begin{thebibliography}{99}

\bibitem{Khoze:2010hg}
V.~A.~Khoze, J.~W.~L\"ams\"a, R.~Orava and M.~G.~Ryskin,
arXiv:hep-ph/1007.3721.

\bibitem{Berestetskii}
V.~B.~Berestetskii, E.~M.~Lifshitz, L.~P.~ Pitaevskii, 
'Relativistic Quantum Theory, Series: Course of Theoretical Physics' , v. 4, Pergamon Press, 1979.

\bibitem{Buras:1973km}
  A.~J.~Buras and J.~Dias de Deus,
  Nucl.\ Phys.\  B {\bf 71} (1974) 481.

\bibitem{Ryskin:2009tj}
  M.~G.~Ryskin, A.~D.~Martin and V.~A.~Khoze,
  Eur.\ Phys.\ J.\  C {\bf 60} (2009) 249
  [arXiv:0812.2407 [hep-ph]];
  
  E.~Gotsman, E.~Levin, U.~Maor and J.~S.~Miller,
  Eur.\ Phys.\ J.\  C {\bf 57} (2008) 689
  [arXiv:0805.2799 [hep-ph]];
  
  S.~Ostapchenko,
  Phys.\ Rev.\  D {\bf 81} (2010) 114028
  [arXiv:1003.0196 [hep-ph]].
  
\bibitem{Fiore:2008tp}
  R.~Fiore, L.~L.~Jenkovszky, R.~Orava, E.~Predazzi, A.~Prokudin and O.~Selyugin,
  Int.\ J.\ Mod.\ Phys.\  A {\bf 24} (2009) 2551
  [arXiv:0810.2902 [hep-ph]].
  
    
 \bibitem{bfk}
 V.~N.~Baier, V.~S.~Fadin and V.~A.~Khoze,
Sov.\ Phys.\ JETP {\bf 23} (1966) 1073
[Zh.\ Eksp.\ Teor.\ Fiz.\  {\bf 50} (1966) 1611].

\bibitem{nov} 
~P.~I.~Golubnichii {\it et al.}
Atomnaya Energiya, {\bf 22} (1966) 168;
Commun. Intern. Symp. Electron and positron storage ring, P.U.F, Saclay
(1966) V.3.1.

\bibitem{ors} 
J.~E.~Augustin {\it et al.},
  Nucl.\ Instrum.\ Meth.\  {\bf 97} (1971) 497. 

\bibitem{Grachov1a}
CMS Collaboration, O.~A.~Grachov et al 
2009 J. Phys.: Conf. Ser. {\bf 160} 012059;

O.~A.~Grachov et al 
AIPConf.Proc.867:258-265, 2006
[arXiv:nucl-ex/0608052v2].

\bibitem{pythia}
PYTHIA 6.2: T.Sj\"ostrand et al,
http://home.thep.lu.se/~torbjorn/PYTHIA 6.2.html 
(Default parameters have been used in the simulations).

\bibitem{geant}
GEANT: S.Agostinelli et al., 
Nucl. Inst. Methods 506 (2003) 250; 
IEEE Trans. Nucl. Sci. 53 (2006) 270; Brun et al., 
GEANT3 Reference Manuel, DD/EE/84-1, CERN 1987.

\bibitem{Khoze:2000db}
  V.~A.~Khoze, A.~D.~Martin, R.~Orava and M.~G.~Ryskin,
  Eur.\ Phys.\ J.\  C {\bf 19} (2001) 313
  [arXiv:hep-ph/0010163].
  
 \bibitem{fsc}
A.~J.~Bell {\it et al.},
Physics and beam monitoring with forward shower counters (FSC) in CMS, CMS NOTE 2010/015; \\ 
M. Albrow, Albert De Roeck, V.A. Khoze, J.W. L\"ams\"a, E. Norbeck, Y. Onel, Risto Orava, and M.G. Ryskin, 
Forward physics with the rapidity gaps at the LHC, JINST 4 : P1001(2009), arXiv:0811.0120v3 [hep-ex] (2009).
  
\end{thebibliography}
\end{document}